\documentclass[12pt]{myels}
\usepackage{epsfig}
\newcommand{\nn}{\nonumber}
\newcommand{\raw}{\rightarrow}

\newcommand{\be}{\begin{equation}}
\newcommand{\ee}{\end{equation}}
\newcommand{\bea}{\begin{eqnarray}}
\newcommand{\eea}{\end{eqnarray}}

\newcommand{\evb}{ {\rm eV$^2$} }

\newcommand{\PCPV}{
\begin{picture}(22,10)
\put(8,-2){\line(2,1){12}}
\put(0,0){$P_{CP}$}
\end{picture}}
\newcommand{\PCPC}{
\begin{picture}(22,10)
\put(0,0){$P_{CP}$}
\end{picture}}
\begin{document}



\begin{frontmatter}
\begin{center}

\title{Update on four-family neutrino oscillations at $\nu$-Factory}




\begin{center}
\author[a]{A. \snm Donini}\footnote{E-mail donini@daniel.ft.uam.es}, 
\author[a]{M.B. \snm Gavela}\footnote{E-mail gavela@garuda.ft.uam.es}, 
\author[b]{P. \snm Hern\'andez}\footnote{On leave from Departamento de 
F\'{\i}sica Te\'orica, Universidad de Valencia. 
E-mail Pilar.Hernandez@cern.ch} and 
\author[c]{S. \snm Rigolin}\footnote{E-mail srigolin@umich.edu}
\address[a]{Departamento de F\'{\i}sica Te\'orica C-XI, Facultad de Ciencias, 
Universidad Aut\'onoma de Madrid, \cty Cantoblanco, Madrid 28049, \cny Spain}
\address[b]{CERN, \cty 1211 Geneve 23, \cny Switzerland}
\address[c]{Department of Physics, University of Michigan, 
\cty Ann Arbor, MI 48015, \cny USA}
\end{center}



\begin{abstract}
The prospects of measuring the leptonic angles and CP-odd phases at a 
{\em neutrino factory} are discussed in the scenario of three active plus 
one sterile neutrino, motivated by the $\nu_\mu \raw \nu_e$ LSND signal.
Sensitivity to the MNS mixing matrix angles at $O (10^{-3})$
is achieved with a 1 T detector at $\sim 1$ km from the source, whereas
leptonic CP-violation can be measured through tau detection with
a 1 kT detector at $\sim 100$ km.
\end{abstract}



\end{center}
\end{frontmatter}


\setcounter{footnote}{0}

%
\section{Introduction}
%
%

Indications in favour of neutrino oscillations have been 
obtained both in solar neutrino 
\cite{Cleveland:1998nv,Fukuda:1996sz,Hampel:1999xg,Abdurashitov:1999zd,Suzuki:1999cy}
and atmospheric neutrino 
\cite{Fukuda:1994mc,Becker-Szendy:1995vr,Fukuda:1999ah,Allison:1999ms,Ambrosio:1998wu} 
experiments.
The atmospheric neutrino data require 
$\Delta m_{atm}^2 \sim (2 - 5) \times 10^{-3}$ \evb \cite{nakamura}
whereas the solar neutrino data prefer $\Delta m_{sun}^2 \sim 10^{-10} - 10^{-4}$ 
\evb, depending on the particular solution to the solar neutrino deficit. 
The LSND data \cite{Athanassopoulos:1998pv} would indicate 
a $\nu_\mu \to \nu_e$ oscillation with a third, very distinct, 
neutrino mass difference: $\Delta m_{LSND}^2 \sim 0.3 - 1$ \evb. 
To explain the present ensemble of data four different 
light neutrino species are needed. The new light neutrino is denoted 
as sterile \cite{Pontecorvo:1968fh}, since it must be an electroweak singlet
to comply with the strong bounds on the $Z^0$ invisible decay width \cite{LEPnu}. 

It has been shown in \cite{Okada:1997kw,Bilenkii:1998rw,Barger:1998bn,Bilenkii:1999ny} 
that the combined analysis of solar, atmospheric and LSND experiments tends 
to exclude a spectrum with three almost degenerate neutrinos and an isolated one. 
It is customary then to consider the following scheme: the lighter pair (1-2) 
is separated by the solar mass difference and the heavier pair (3-4) by the 
atmospheric one\footnote{The alternative option, 
with a small separation in the heavier pair and a larger one 
for the lighter pair, amounts to reversing the sign of the LSND 
mass difference. Matter effects could be important in order to
establish the physical spectrum, much as in the case of three-family
neutrino mixing 
\cite{Freund:2000gy,Cervera:2000kp,Barger:2000yn,golden2}.}. 
We work in the convention $\Delta m^2_{12} = \Delta m^2_{sol}$, 
with the solar deficit assigned mainly to $\nu_s-\nu_e$ oscillations, 
and $\Delta m^2_{34} = \Delta m^2_{atm}$,
with the atmospheric anomaly mainly due to $\nu_\mu-\nu_\tau$ oscillations. 
Our flavour basis is $\nu_\alpha = ( \nu_s, \nu_e, \nu_\mu, \nu_\tau )$.
The alternative possibility of identifying the atmospheric anomaly as a 
$\nu_\mu-\nu_s$ oscillation is excluded at 99 \% C.L. by SK data \cite{nakamura}
(see also \cite{Giunti:2000ye,Yasuda:2000dc} at this conference).
We choose the following parametrization:
\be
U = U_{14} (\theta_{14}) U_{13} (\theta_{13}) U_{24} (\theta_{24}) 
    U_{23} (\theta_{23},\delta_3) U_{34} (\theta_{34},\delta_2) 
    U_{12} (\theta_{12},\delta_1)
\label{ourpar}
\ee
and consider two approximations \cite{Donini:1999jc,Donini:1999he}, 
\begin{enumerate}
\item
$\Delta m_{12}^2 = \Delta m_{34}^2 = 0$, ``one mass scale dominance'' (or 
minimal) scheme;
\item
$\Delta m_{12}^2 = 0$, ``two mass scale dominance'' (or next-to-minimal) 
scheme.  
\end{enumerate}
If the eigenstates $i$ and $j$ are degenerate and the matrix $U_{ij}$ is located 
to the right in eq.~(\ref{ourpar}), the angle $\theta_{ij}$ becomes automatically 
unphysical and drops from the oscillation probability expressions. 
When a different ordering is taken no angle drops from the oscillation probabilities. 
A redefinition of the rest of the parameters is necessary  
to illustrate the remaining reduced parameter space in a transparent way.  
Finally, we associate the phases to the rightmost rotation matrix: 
in this way, the right number of independent physical parameters automatically 
appears in the transition probabilities whenever $\Delta m^2_{12}$ 
or both $\Delta m^2_{12}, \Delta m^2_{34}$ are neglected. 

\section{Sensitivity reach of the {\em neutrino factory} }
\label{sect:sensi}

We concentrate now on the sensitivity to the different angles in
the ``one mass scale'' approximation. 
Four rotation angles ($\theta_{13}$, $\theta_{14}$, $\theta_{23}$ and 
$\theta_{24}$) and one complex phase ($\delta_3$) remain. The two rotation 
angles that have become unphysical are already tested at solar ($\theta_{12}$ 
in our parametrization) and atmospheric ($\theta_{34}$) neutrino experiments. 
The remaining four can be studied at the {\em neutrino factory} with high 
precision. 

We consider a muon beam of $E_\mu = 20$ GeV, 
for a beam intensity of $2 \times 10^{19}$ and
$2 \times 10^{20}$ useful $\mu^-$ per year. 
The large LSND mass difference, $\Delta m_{23}^2 \simeq 1 {\rm eV}^2$, 
calls for a SBL experiment, rather than for a LBL one. 
We consider therefore a 1 T detector located at $\sim 1$ km distance 
from the neutrino source, resulting in $N_{CC} \simeq 10^7$ charged leptons detected.
An efficiency of $\epsilon_\mu = 0.5, \epsilon_\tau = 0.35$ for $\mu$ and $\tau$ 
detection respectively, and a background contamination 
at the level of $10^{-5} N_{CC}$ events are included. 

The existing experimental data impose some constraints on the parameter 
space. Bugey and Chooz \cite{Declais:1995su,Apollonio:1999ae}
measure the oscillation probability 
\be
P(\bar{\nu}_e \to \bar{\nu}_e) = 1
 - 4 c^2_{23} c^2_{24} ( s^2_{24} + s^2_{23} c^2_{24} ) 
       \sin^2 \left ( \frac{\Delta_{23}}{2} \right )
\ee
The allowed mass range for the LSND signal 
of $\nu_e \to \nu_\mu$ transitions provides the constraint 
$10^{-3} \leq c^2_{13} c^2_{24} \sin^2 2 \theta_{23} \leq 10^{-2}$, 
which fits nicely with the Chooz constraint to point towards small $s_{23}^2$ values.
We choose to be ``conservative'', or even ``pessimistic'', in order to 
illustrate the potential of the {\em neutrino factory}. In the numerical 
computations below we will make the assumption that all angles crossing 
the large LSND gap, $\theta_{13}$, $\theta_{14}$, $\theta_{23}$ and 
$\theta_{24}$ are small.

For small values of $\theta_{23}$, the best sensitivity to $\theta_{13}$ 
is given by the $\mu^-$ disappearance channel;
the $\mu^+$ appearance channel, in turn, is more sensitive to $\theta_{23}$;
eventually, the $\tau^-$ appearance channel is best suited to measure
$\theta_{14}$ and $\theta_{24}$. The corresponding transition probabilities
are:
\bea
\PCPC(\nu_\mu \to \nu_\mu) &=& 1
 - 4 c^2_{13} c^2_{23} ( s^2_{23} + s^2_{13} c^2_{23} ) \
       \sin^2 \left ( \frac{\Delta_{23}}{2} \right ) \ , 
\label{cpeven2} \\
\PCPC(\nu_e \to \nu_\mu) &=& 4 c^2_{13} c^2_{24} c^2_{23} s^2_{23} \
\sin^2 \left ( \frac{\Delta_{23}}{2} \right ) \ , 
\label{cpeven1} \\
\PCPC(\nu_\mu \to \nu_\tau)  &=&  4 c^2_{13} c^2_{23} 
\left [ ( s^2_{13} s^2_{14} c^2_{23} + c^2_{14} s^2_{23} s^2_{24} ) 
\right . \nn \\
& + & \left .  2 c_{14} s_{14} c_{23} s_{23} s_{13} s_{24} 
\cos \delta_3 \right ] \
\sin^2 \left ( \frac{\Delta_{23}}{2} \right ) \, .
\label{cpeven4}
\eea
The $\tau^+$ appearance channel is worse than the $\tau^-$ one due to 
cancellations between the two terms in the probability (see 
\cite{Donini:1999jc,Donini:1999he} for details). 

Notice that the physical phase appears in $\PCPC(\nu_\mu \to \nu_\tau)$ 
in a cosine dependence. Actually, no CP-odd observable can be build out 
of the oscillation probabilities in this approximation in spite of the existence 
of a physical CP-odd phase in the mixing matrix. 

\begin{figure}[t]
\hspace{3cm}
\vspace{0.1cm}
\epsfig{figure=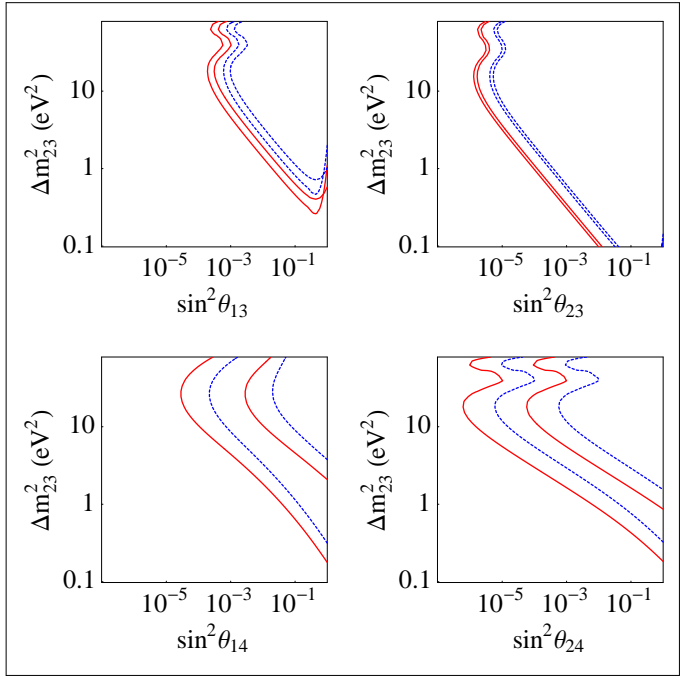,height=10cm,angle=0}
\caption{\it{
Sensitivity reach in the $s^2_{ij} / \Delta m_{23}^2$ plane
for a 1 T detector at $L = 1 km$ from the source. 
Solid (dashed) lines are $N_{\mu} = 2 \times 10^{20}$ ($2 \times 10^{19}$)
useful muons per year; 
(a) reach for $s^2_{13}$ for $\theta_{23} = 10^\circ, 30^\circ$ (from left to right),
$\theta_{14} = \theta_{13}, \theta_{24} = 1^\circ$,
in the $\mu$-disappearance channel;
(b) reach for $s^2_{23}$ for $\theta_{13} = 1^\circ, 10^\circ$ (from right to left),
$\theta_{14} = \theta_{23}, \theta_{24} = 1^\circ$,
in the $\mu$-appearance channel;
(c) reach for $s^2_{14}$ for $\theta_{13} = 1^\circ, 10^\circ$ (from right to left),
$\theta_{23} = 10^\circ, \theta_{24} = 1^\circ$,
in the $\tau^-$-appearance channel;
(d) reach for $s^2_{24}$ for $\theta_{23} = 1^\circ, 10^\circ$ (from right to left),
$\theta_{13} = \theta_{14} = 10^\circ$,
in the $\tau^-$-appearance channel.
}} 
\label{fig:all_angles}
\end{figure}

In Fig. \ref{fig:all_angles} we present
the sensitivity to $\theta_{13}$ through the $\mu$-disappearance channel 
for different values of $\theta_{23}$ (a); 
to $\theta_{23}$ through the $\mu$-appearance channel 
for different values of $\theta_{13}$ (b);
to $\theta_{14}$ for different values of $\theta_{13}$ (c) 
and to $\theta_{24}$ for different values of $\theta_{23}$ (d)
through the $\tau^-$ appearance channel.
In all cases, we see that the $\nu$-factory can easily measure
$\sin^2 \theta_{ij}$ down to $10^{-3}-10^{-2}$
for not too tiny mass difference, both for
$N_\mu = 2 \times 10^{19}, 2 \times 10^{20}$ useful muons per year. 

\section{CP Violation with four neutrino species}
\label{sect:4CPviolation}

As in the standard three-family scenario, to observe CP-odd effects in oscillations 
it is necessary to have both physical CP-odd phases and at least two non-vanishing 
mass differences. The next-to minimal or ``two mass scale dominance'' scheme, 
described at the beginning of this section, is thus suitable.

CP-odd effects are observable in ``appearance'' channels, while 
``disappearance'' ones are only sensitive to the CP-even part. 
In contrast with the three-neutrino
case, the solar suppression (see \cite{DeRujula:1999hd}) is now replaced by a less 
severe atmospheric suppression. CP-violating effects are expected to 
be at least one order of magnitude larger than in the standard case, 
because they are not suppressed by the solar mass-difference, $\Delta m^2_{12}$.

As explained above, the parameter space consists of five angles and two 
CP-odd phases. Staying in the ``conservative'' assumption of small 
$\theta_{13},\theta_{14}, \theta_{23}, \theta_{24}$, we compare two democratic 
scenarios, in which all of these angles are taken to be small and of the same order:
\begin{enumerate}
\item Set 1: $\theta_{34} = 45^\circ$, $\theta_{ij} = 5^\circ$ 
      and $\Delta m^2_{atm} = 2.8 \times 10^{-3}$ eV$^2$
      for $\Delta m^2_{LSND} = 0.3$ eV$^2$;
\item Set 2: $\theta_{34} = 45^\circ$, $\theta_{ij} = 2^\circ$ and 
      $\Delta m^2_{atm} = 2.8 \times 10^{-3}$ eV$^2$ for 
      $\Delta m^2_{LSND} = 1$ eV$^2$.
\end{enumerate}
For illustration we consider in what follows a 1 kT detector located at 
$O(10)$ km distance from the neutrino source. 

The CP-odd $\nu_\mu \to \nu_\tau$ probability is (to the leading order in 
$\Delta m^2_{atm}$, i.e. $\Delta m^2_{34}$ in our parametrization)
\bea
\PCPV(\nu_\mu \to \nu_\tau) & = & 
      8 c^2_{13} c^2_{23} c_{24} c_{34} s_{34} \left [ 
      c_{14} c_{23} s_{13} s_{14} \ \sin \delta_2 \ + 
      c^2_{14} s_{23} s_{24} \ \sin (\delta_2 + \delta_3) \right ] \times \nn \\ 
&   & \qquad \left( {{ \Delta_{34}}\over{2} } \right )  
      \sin^2 \left( \frac{\Delta_{23} }{2} \right ) .
\label{cpodd2}
\eea
We define as in \cite{DeRujula:1999hd,Cervera:2000kp} 
the neutrino-energy integrated asymmetry: 
\be
{\bar A}^{CP}_{\mu \tau} (\delta) = \frac{ \{ N[\tau^+]/N_o[\mu^+] \} 
                                         - \{ N[\tau^-]/N_o[\mu^-] \} }{
                                           \{ N[\tau^+]/N_o[\mu^+] \} 
                                         + \{ N[\tau^-]/N_o[\mu^-] \} }\; ;
\label{intasy}
\ee
with $N[\tau^\pm]$ the measured number of taus, and 
$N_o[\mu^\pm]$ the expected number of muons 
charged current interactions in the absence of oscillations.
In order to quantify the significance of the signal, we compare the 
value of the integrated asymmetry with its error, in which we include the 
statistical error and a conservative background estimate at
the level of $10^{-5}$. The size of the CP-asymmetries is very different 
for $\mu$- and $\tau$-appearance channels. 
For instance for Set 2, they turn out to be small 
in $\nu_e-\nu_\mu$ oscillations, ranging from the per mil level to a few percent. 
In contrast, in $\nu_\mu-\nu_\tau$ oscillations they attain much larger values 
of about $50\%-90\%$. These larger factors follow from the fact that the CP-even 
transition probability $P_{CP}(\nu_\mu \to \nu_\tau)$ is smaller than 
$P_{CP}(\nu_e \to \nu_\mu)$, due to a  stronger suppression in small mixing angles
(see eqs. (\ref{cpeven1},\ref{cpeven4}).
Notice that the opposite happens in the 3-species case.

\begin{figure}[t]
\vspace{0.1cm}
\begin{tabular}{cc}
\epsfig{figure=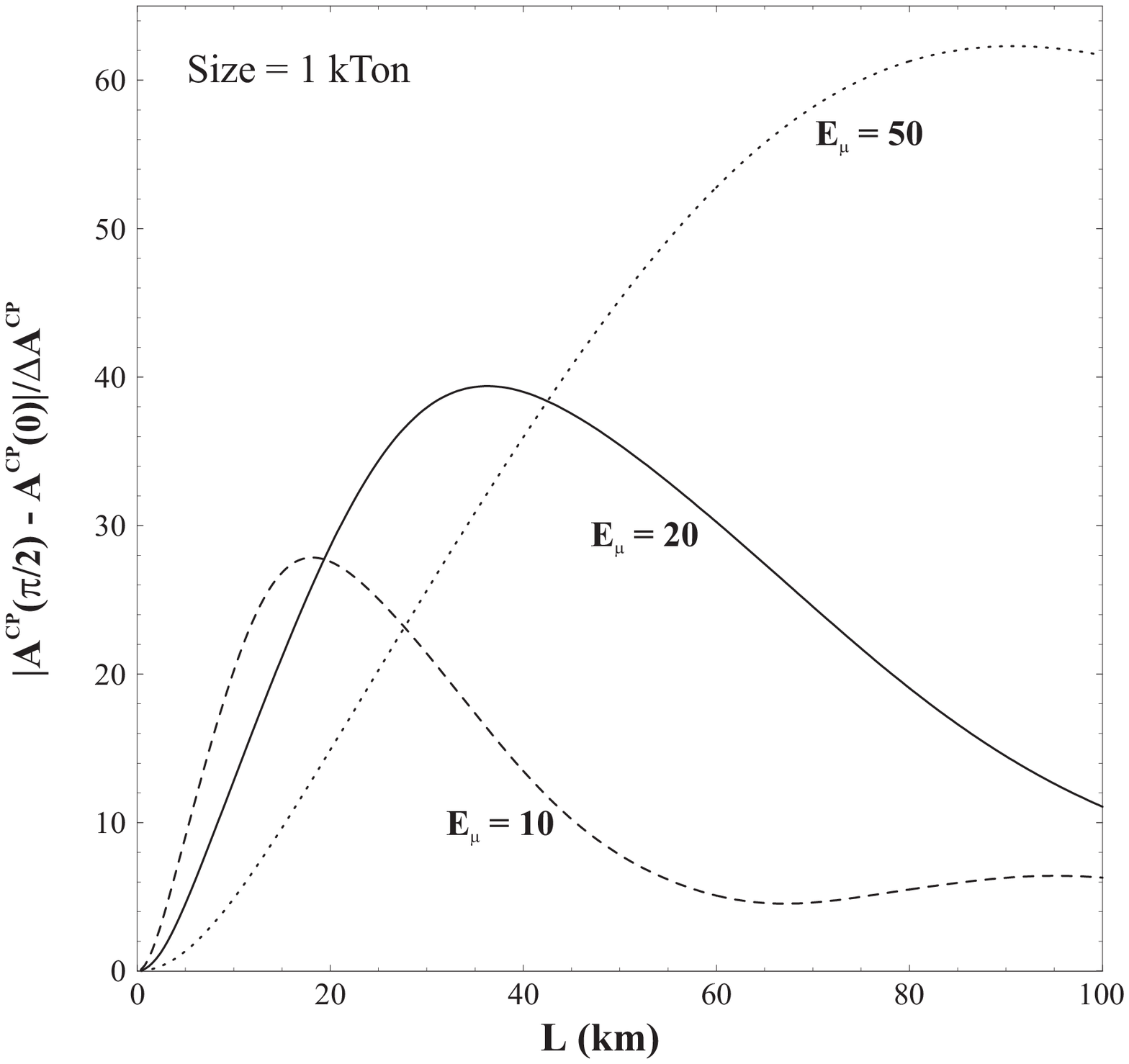,height=7.5cm,angle=0} & 
\epsfig{figure=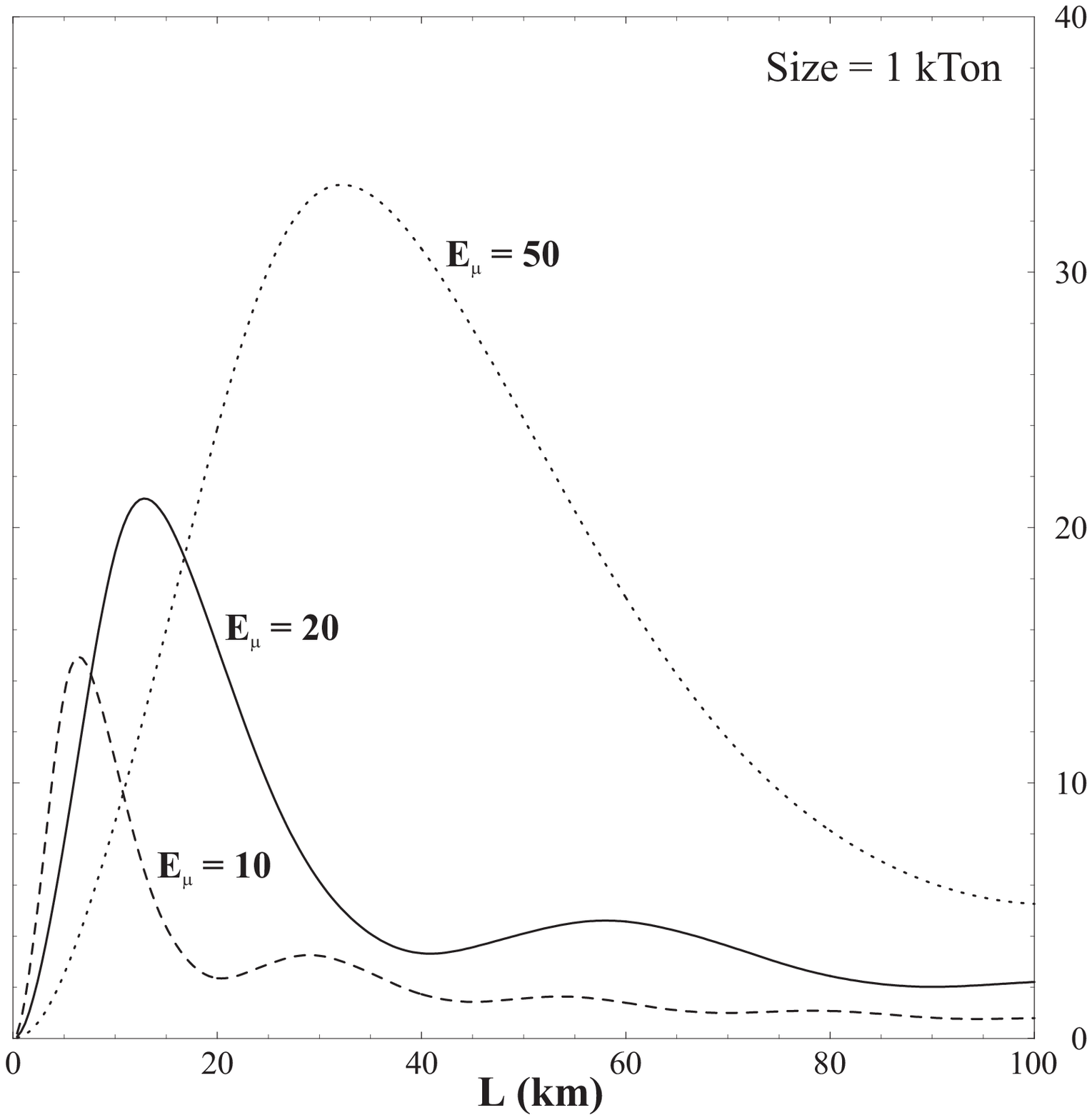,height=7.5cm,angle=0} 
\end{tabular}
\caption{\it{
Signal over statistical uncertainty for CP violation,
in the $\nu_\mu \to \nu_\tau$ channel, for the two sets of parameters described 
in the text (Set 1 on the left and Set 2 on the right). We consider a $1$ kT 
detector and $2 \times 10^{20}$ useful muons/year.}} 
\label{CPmtfig12}
\end{figure}

In Fig.~\ref{CPmtfig12} we show the signal over noise ratio in $\nu_\mu 
\to \nu_\tau$ versus $\bar\nu_\mu \to \bar\nu_\tau$ oscillations as a 
function of the distance. Matter effects, although negligible
(see \cite{Kalliomaki:1999ii}), have been included,
as well as the exact formulae for the probabilities.
For the scenario and distances discussed here, the scaling laws are analogous 
to those derived for three neutrino species in vacuum. 
The maxima of the curves move towards larger distances when the energy of the 
muon beam is increased, or the assumed LSND mass difference is decreased. 
Moreover, increasing the energy enhances the significance of the effect at the 
maximum. 
At $E_\mu = 50$ GeV, 60 standard deviation (sd) signals are attainable at 
around 100 km for the values in Set 1, and 30 sd around 40 km for Set 2, 
levelling off at larger distances and finally diminishing when $E_\nu/L$ 
approaches the atmospheric range.

\end{document}